\documentclass[10pt,journal]{IEEEtran} 

\usepackage[utf8]{inputenc}
\usepackage[T1]{fontenc}
\usepackage{amssymb,amsmath}
\usepackage{cite}
\usepackage{psfrag}
\usepackage{url}
\usepackage[absolute,overlay]{textpos}
\DeclareMathOperator{\Tr}{Tr}
\usepackage{pgf}
\usepackage[export]{adjustbox}
\usepackage{bm}
\usepackage{bbm}
\usepackage{booktabs}
\usepackage{url}
\usepackage{upgreek}
\usepackage{verbatimbox}
\usepackage{algorithm}
\usepackage{algorithmic}

\usepackage{amsthm}

\begin{document}
	
	\title{A Low-Complexity Beamforming Design for Multiuser Wireless Energy Transfer}	
	
	\author{Onel L. A. López, 
		Francisco~A.~Monteiro, 
		Hirley Alves, 
		Rui Zhang, 
		and Matti Latva-aho, 
		\thanks{Onel López, Hirley Alves, and Matti Latva-aho are  with  the Centre for Wireless Communications University of Oulu, Finland, e-mails: \{Onel.AlcarazLopez, Hirley.Alves, Matti.Latva-aho\}@oulu.fi.}
		\thanks{F. A. Monteiro is with Instituto de Telecomunicações, and ISCTE - Instituto Universitário de Lisboa, Portugal, e-mail: francisco.monteiro@lx.it.pt.}
		\thanks{Rui Zhang is with the National University of Singapore, Singapore, e-mail: elezhang@nus.edu.sg}
		\thanks{This work is supported by Academy of Finland (Aka) (Grants n.307492, n.318927 (6Genesis Flagship), n.319008 (EE-IoT)), FCT (Foundation for Science and Technology) and Instituto de Telecomunica\c{c}\~{o}es through national funds, and when applicable co-funded EU funds, under the project UIDB/EEA/50008/2020.}
		\thanks{\copyright 2020 IEEE. This paper has been accepted for publication in IEEE Wireless Communications Letters. Personal use of this material is permitted. Permission from IEEE must be obtained for all other uses, in any current or future media, including reprinting/republishing this material for advertising or promotional purposes, creating new collective works, for resale or redistribution to servers or lists, or reuse of any copyrighted component of this work in other works.}
	} 
	
	\maketitle
	
	\begin{abstract}
		Wireless energy transfer (WET) is a green enabler of low-power Internet of Things (IoT). Therein, traditional optimization schemes relying on full channel state information (CSI) are often too costly to implement due to excessive  energy consumption and high processing complexity. This letter proposes a simple, yet effective, energy beamforming scheme that allows a multi-antenna power beacon (PB) to fairly power a set of IoT devices by only relying on the first-order statistics of the channels. In addition to low complexity, the proposed scheme
		performs favorably as compared to benchmarking schemes and its performance improves as the number of PB's antennas increases. Finally, it is shown that further performance improvement can be achieved through proper angular rotations of the PB.
	\end{abstract}
	\section{Introduction}\label{intro}
	Wireless energy transfer (WET) technology is widely recognized as a green enabler of low-power Internet of Things (IoT) since it realizes \cite{Lopez.2019}: i) battery charging without physical connections, which simplifies servicing and maintenance; and ii) form factor reduction and durability increase of the end devices. In fact, IoT industry is already strongly betting on this promising technology, proof of which is the variety of emerging enterprises with a large portfolio of WET solutions, e.g., Powercast, TransferFi and Ossia\footnote{See \url{https://www.powercastco.com//},  \url{https://www.transferfi.com/} and \url{https://www.ossia.com}}. 
	
	Over the past few years, the research community has been analyzing and optimizing WET-enabled communication systems. However, emphasis has been given to the information communication aspects rather than to the WET building block itself, which in fact causes the performance bottleneck in practical applications since 
	WET of long duration is often required  so that the energy harvesting (EH) devices can harvest sufficient amount of energy for operation and communication.
	Nevertheless, there are works dedicated to WET in the literature. For instance, the authors of \cite{Zeng.2015} designed a channel state information (CSI) acquisition method for a point-to-point multiple-input multiple-output (MIMO) WET system by exploiting the channel reciprocity, such that full benefits from energy beamforming (EB) can be practically obtained. Such results were extended to frequency-selective channels but under a multiple-output single-output (MISO) scenario in \cite{ZengZhang.2015}. 
	Additionally, the problem of  power beacons (PBs) deployment optimization is addressed in \cite{Dai.2018}, 
	the feasibility of WET using massive MIMO has been corroborated in \cite{Kashyap.2016}, while in \cite{Chu.2018} authors exploit \textit{energy trading} mechanisms in a scenario where the PB and EH devices belong to different operators. Also, a method relying on multiple dumb antennas transmitting phase-shifted signals  to induce fast fluctuations on a slow-fading wireless channel and attain transmit diversity is proposed in \cite{Clerckx.2018}. This scheme has the additional advantage of being CSI-free, i.e., the CSI is not required at the transmitter/receiver. Therefore, particularly beneficial for radio-frequency (RF) EH networks since CSI acquisition consumes energy of ultra-low power devices and thus may not be affordable. 
	The CSI acquisition problem takes on larger dimensions as the network densifies since the gains from CSI-based EB quickly decrease  as the number of EH devices grows larger. Consequently, efficient CSI-limited/free schemes are required for enabling low-power massive IoT \cite{Lopez.2019}. Based on this ground rule, the authors in \cite{LopezAlves.2019,LopezMontejo.2020} have proposed and optimized several multi-antenna CSI-free WET solutions to improve the statistics of the RF energy availability at the input of the EH circuitry of a massive set of energy harvesters. However, their full gains  are obtained only in truly massive setups, while for a moderate number of devices their performance is not promising. 
	
	In this letter, a low-complexity EB scheme is proposed for a multi-antenna PB to wirelessly power a set of single-antenna EH devices. The main contributions are three-fold: i) different from the existing works, this letter addresses the problem of powering devices with fairness while using only channels' first-order statistics; ii) a simple, yet effective, EB scheme is proposed, which attains near-optimum performance as the channels become more deterministic; iii) it is demonstrated that the multiuser performance improves as the number of transmit antennas increases, while further performance improvement can be obtained via proper angular rotation. 
	
	\textbf{Notation:} boldface lowercase letters denote column vectors, while boldface uppercase letters denote matrices. For instance, $\mathbf{x}=\{x_i\}$, where $x_i$ is the $i$-th element of vector $\mathbf{x}$, while
	$\mathbf{X} = \{X_{i,j}\}$, where $X_{i,j}$ is the $i-$th row $j-$th column element of matrix $\mathbf{X}$.
	$\mathbf{1}$ denotes a vector of ones, $\mathbf{I}$ denotes the identity matrix, and $\mathrm{diag}(\mathbf{x})$ is a diagonal matrix with the main diagonal from entries of $\mathbf{x}$.  $||\mathbf{x}||$ denotes the Euclidean norm of $\mathbf{x}$, while $(\cdot)^T$, $(\cdot)^H$, $\Tr(\cdot)$, and $|\cdot|$ denote the transpose, Hermitian transpose, trace, and absolute value operations, respectively. The curled inequality symbol $\succeq$ is used to denote generalized inequality: between vectors, it represents component-wise inequality; between symmetric matrices, it represents matrix inequality. Also, $\inf\{\cdot\}$, $\min\{\cdot\}$, $\max\{\cdot\}$ and $\mathcal{O}(\cdot)$ are the infimum,  minimum, maximum and big-O notations, respectively. $\mathbb{C}$ is the set of complex numbers, and $\mathbbm{i}=\sqrt{-1}$ is the imaginary unit. Finally, $\mathbf{x}\sim\mathcal{CN}(\mathbf{m},\mathbf{R})$ is a circularly-symmetric complex Gaussian random vector with mean vector $\mathbf{m}$ and covariance matrix $\mathbf{R}$, and $\mathbb{E}[\ \!\cdot\! \ ]$ denotes the statistical expectation. 
	\section{System Model}\label{system}
\begin{figure}[t!]
		\centering
		\includegraphics[width=0.3\textwidth]{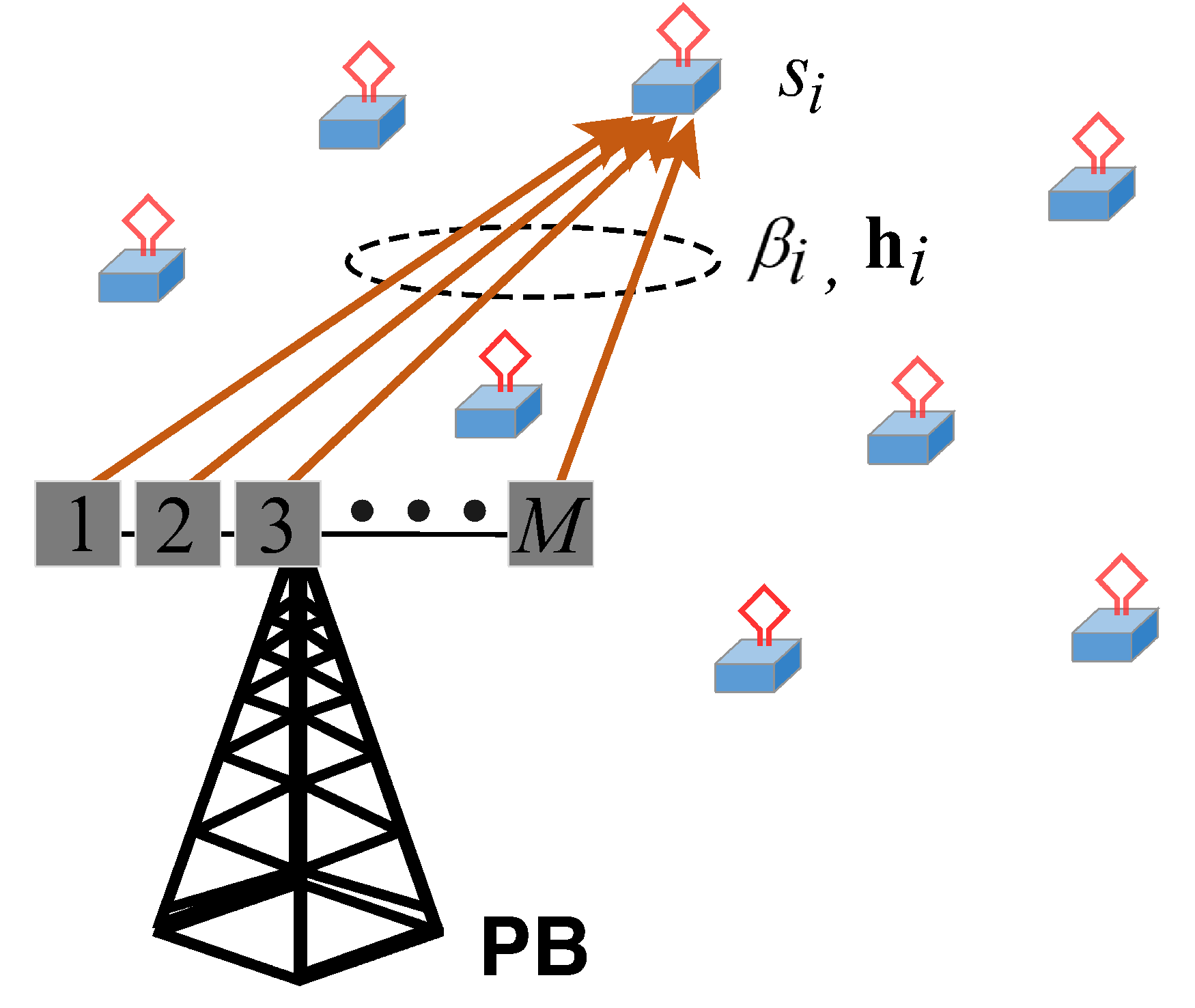}
		\caption{Illustration of the system model. A PB equipped with $M$ antennas wirelessly powers a set of $N\le M$ IoT devices.}	
		\label{Fig1}
	\end{figure}

	Consider the scenario illustrated in Fig.~\ref{Fig1} where a PB equipped with $M$ antennas transfers energy via RF to a set $\{s_i\}$ of $N\le M$ single-antenna IoT devices located nearby. Quasi-static channels are assumed, with fading remaining constant over a transmission block, and changing from block to block with unknown distribution.
	The power-normalized channel vector between the PB's antennas and $s_i$ is denoted as $\mathbf{h}_i=\mathbf{\bar{h}}_i+\mathbf{\hat{h}}_i\in \mathbb{C}^{M\times 1}$, where $\mathbf{\bar{h}}_i$ is the deterministic component and $\mathbf{\hat{h}}_i$ is the zero-mean random component with covariance  $\mathbf{R}_i\!=\!\mathbb{E}[\mathbf{\hat{h}}_i\mathbf{\hat{h}}_i^H]$. Moreover, $\beta_i$ denotes the average power gain of the channel between the PB and $s_i$. 
	
	The PB transmits $K\le N$ complex signals $\{x_k\}$ such that the received RF signal available at each device $s_i$ (with the noise ignored and normalized transmit power) is given by
	\begin{align}
	y_i = \sqrt{\beta_i}\mathbf{h}_i^T\sum_{k=1}^K\mathbf{w}_kx_k, \ \ i=1 \cdots N,
	\end{align}
	where $\mathbf{w}_k\in\mathbb{C}^{M\times 1}$ represents the precoding vector associated to the $x_k$, thus $\sum_{k=1}^K||\mathbf{w}_k||^2=1$. Signals are assumed  independent and normalized such that $\mathbb{E}[x_k^Hx_k]=1$ and $\mathbb{E}[x_k^Hx_j]=0, \forall k\ne j$. As such, the RF energy (normalized by unit time) available at each $s_i$ is given by
	\begin{align}
	E_i &= \mathbb{E}_{x}[y_i^Hy_i]=\beta_i\mathbb{E}_{x}\bigg[\Big(\sum_{k=1}^K\mathbf{h}_i^T\mathbf{w}_kx_k\Big)^{H}\Big(\sum_{k=1}^K\mathbf{h}_i^T\mathbf{w}_kx_k\Big)\bigg]\nonumber\\ 
	&=\beta_i\sum_{k=1}^K\big|\mathbf{h}_i^T\mathbf{w}_k\big|^2\mathbb{E}_x[x_k^Hx_k]=\beta_i\sum_{k=1}^K\big|\mathbf{h}_i^T\mathbf{w}_k\big|^2.\label{Ei}
	\end{align}
	\section{Problem Formulation}\label{problem}
	The goal is  to maximize the amount of energy harvested  per device in a fair manner, which is formulated as the following optimization problem
	\begin{subequations}\label{P}
		\begin{alignat}{2}
		\mathbf{P1:}\qquad &\underset{\{\mathbf{w}^{(j)}\},\ \forall j}{\mathrm{maximize}}       &\qquad & 
		\inf_{i=1,\cdots,N}\   \{E_i\} \label{P:a}\\
		&\text{subject to} & &  \sum_{k=1}^{K}||\mathbf{w}_k||^2\leq 1. \label{P:b}
		\end{alignat}	
	\end{subequations}
	Notice that the constraint in \eqref{P:b} is convex and it is related to the total transmit power, while the objective function in \eqref{P:a} is not concave, therefore the problem is not convex. However, it can  still be  optimally solved by rewriting it as a semi-definite programming (SDP) problem \cite{Thudugalage.2016} as shown next.
	
	First, define $\xi \triangleq \text{inf} \{E_i\}$, while $E_i$ in \eqref{Ei} can be rewritten as $E_i=\beta_i\sum_{k=1}^K\mathbf{h}_i^H\mathbf{w}_k\mathbf{w}_k^H\mathbf{h}_i=\beta_i\Tr(\mathbf{W}\mathbf{H}_i)$, where $\mathbf{W}=\sum_{k=1}^K\mathbf{w}_k\mathbf{w}_k^H$ and $\mathbf{H}_i=\mathbf{h}_i\mathbf{h}_i^H$. Second, notice that $\mathbf{W}$ is a Hermitian matrix with maximum rank $M$ that can be found by solving the SDP:
	\begin{subequations}\label{P12}
		\begin{alignat}{2}
		\mathbf{P2:}\ \ &\underset{\mathbf{W}\in\mathbb{C}^{M\times M},\ \xi}{\mathrm{minimize}}       &\ \ \ & 
		-\xi \label{P12:a}\\ 
		&\text{subject to} & &  \beta_i\Tr(\mathbf{W}\mathbf{H}_i)\ge \xi, \  i\!=\!1,\cdots,N \label{P12:b}\\ 
		& & &\qquad \Tr(\mathbf{W})= 1 \label{P12:c}\\
		& & &\qquad\qquad \mathbf{W}\succeq 0.  \label{P12:d}
		\end{alignat}	
	\end{subequations}
	Notice that constraint  \eqref{P12:c} is equivalent to \eqref{P:b}. After solving $\mathbf{P2}$, the beamforming vectors $\{\mathbf{w}_k\}$, with $K$ equal to the rank of $\mathbf{W}$, can be obtained as  the eigenvectors of $\mathbf{W}$, which is referred to  as \textbf{optimum full-CSI} beamforming.
	\subsection{Statistical Beamforming Design} \label{statistical}
	Notice that for optimally solving  $\mathbf{P1}$ and $\mathbf{P2}$, the instantaneous CSI vectors need to be perfectly known at the PB. However, in practice, not only CSI is imperfect but  devices' cooperation is also required for its acquisition, which consumes their harvested energy. In some cases, the EB gains cannot compensate  the energy consumed during the CSI acquisition, and as a result the net harvested energy of devices becomes negative \cite{Zeng.2015,ZengZhang.2015}.
    In order to mitigate these adverse effects, herein we focus on the average harvested energy optimization. It can be observed that
	\begin{align}
	\mathbb{E}[E_i]&=\mathbb{E}\big[\beta_i\Tr(\mathbf{W}\mathbf{H}_i)\big]\nonumber\\
	&\stackrel{(a)}{=}\!\mathbb{E}\big[\beta_i\Tr(\mathbf{W}\mathbf{\bar{H}}_i)\!+\!2\beta_i\Tr(\mathbf{W}\tilde{\mathbf{H}}_i)\!+\!\beta_i\Tr(\mathbf{W}\mathbf{\hat{H}}_i)\big]\nonumber\nonumber\\
	&\stackrel{(b)}{=}\beta_i\Tr\big(\mathbf{W}(\mathbf{\bar{H}}_i+\mathbf{R}_i)\big),\label{EE}
	\end{align}
	where $(a)$ comes from 
	\begin{align}
	\mathbf{H}_i&=\mathbf{h}_i\mathbf{h}_i^H=(\mathbf{\bar{h}}_i+\mathbf{\hat{h}}_i)(\mathbf{\bar{h}}_i+\mathbf{\hat{h}}_i)^H\nonumber\\
	&=\underbrace{\mathbf{\bar{h}}_i\mathbf{\bar{h}}_i^H}_{\mathbf{\bar{H}}_i}+\underbrace{\mathbf{\bar{h}}_i\mathbf{\hat{h}}_i^H+\mathbf{\hat{h}}_i\mathbf{\bar{h}}_i^H}_{\tilde{\mathbf{H}}_i}+\underbrace{\mathbf{\hat{h}}_i\mathbf{\hat{h}}_i^H}_{\mathbf{\hat{H}}_i},\nonumber
	\end{align}
	and $(b)$ comes from taking the expectation inside the trace, which is a linear operator, and using $\mathbb{E}[\mathbf{\bar{H}}_i]=\mathbf{\bar{H}}_i$, $\mathbb{E}[\tilde{\mathbf{H}}_i]\!=\!\mathbf{0}$, and $\mathbb{E}[\mathbf{\hat{H}}_i]=\mathbf{R}_i$. From \eqref{EE}, it becomes evident that the optimum statistical-CSI beamforming, $\{\mathbf{w}_k^*\}_{\forall k}$, can be obtained by solving $\mathbf{P2}$ but using $\mathbf{\bar{H}}_i+\mathbf{R}_i$ instead of  $\mathbf{H}_i$. Note also that $\mathbb{E}[\mathbf{\bar{H}}_i]=\bar{E}_i+\hat{E}_i$, where $\bar{E}_i=\beta_i\Tr(\mathbf{W}\mathbf{\bar{H}}_i)> 0$ and $\bar{E}_i=\beta_i\Tr(\mathbf{W}\mathbf{R}_i)> 0$ correspond to the average energy associated to the first- and second-order channel statistics, respectively.
	In this letter, we consider only average CSI is available, i.e., only the first-order statistics of the channels, thus only $\{\mathbf{\bar{H}}_i\}$ are assumed to be known. 
	Such average CSI is much less prone to estimation errors and, more importantly, it varies over a much larger time scale and does not require frequent CSI updates. This information is expected to be  beneficial since 
	WET channels are typically  LOS-dominant due to the short distances, and  consequently have strong deterministic components\footnote{CSI acquisition procedures may be further reduced  by exploiting information related to the PB's antenna array architecture and devices' positioning, which influence the line-of-sight (LOS) channel, and consequently the channel's deterministic component, the most.}. Note that the resulting \textbf{optimum average-CSI} beamforming solution is indeed optimal when the channels tend to be fully deterministic, i.e., $\mathbf{\hat{h}}_k\rightarrow \mathbf{0},\ \forall k$. 
	\subsection{Problem Complexity}
	The \textbf{optimum average-CSI} beamforming comes from solving $\mathbf{P2}$ using $\mathbf{\bar{H}}_i$ instead of $\mathbf{H}_i$, thus, optimizing $\inf\{\bar{E}_i\}$, which constitutes a lower bound for the actual average harvested energy since $\mathbb{E}[E_i]>\bar{E}_i$. 
	Now, since $\mathbf{P2}$ is an SDP, interior point methods are adopted to efficiently find its optimal solution. It can be shown that solving $\mathbf{P2}$ requires around $\mathcal{O}(\sqrt{M}\log({1/\varepsilon}))$ iterations, with each iteration requiring at most $\mathcal{O}(M^6+(N+1)M^2)$ arithmetic operations \cite{Ye.2011}, and where $\varepsilon$ represents the solution accuracy attained when the algorithm ends.
	Consequently, the SDP solution becomes computationally costly as the number of antennas at the PB increases.
	To achieve complexity reduction, a low-complexity beamforming design, that attains near-optimum performance as the channels become more deterministic,  is proposed  next as an efficient alternative.
	The proposed scheme can be easily adapted to scenarios including information transmission, e.g., wireless powered communication networks (WPCN), and simultaneous wireless information and power transfer (SWIPT).
	\section{Low-Complexity Beamforming Design}\label{EB}
	The key of the proposed design lies in finding the EB $\{\mathbf{w}_k\}_{\forall k}$ that maximizes $\bar{\xi}$ such that $\bar{E}_i\geq \bar{\xi},\ \forall i$.
	The complexity of the problem can be alleviated by adopting 
	\begin{align}
	\mathbf{w}_k^T= \frac{{\mathbf{\bar{h}}_k}^H}{||\mathbf{\bar{h}}_k||}\sqrt{p_k},\qquad k=1,\cdots,K, \label{wk}
	\end{align}
	with $K=N$. In doing this, the $i-$th signal transmitted from all PB's antennas  arrives at $s_i$ with constructive superposition, in an average sense. In fact, this design is akin to the maximum ratio transmission (MRT) in MISO communications. Notice that 
	$p_k$ represents the power budget for $x_k$ such that $\sum_{i=1}^Np_i=1$, 
	while one should notice that the impact of the signals for other devices  on the RF energy at $s_i$ is not considered for the phases' design. Thus, the RF energy over the deterministic component of channels can be written as
	\begin{align}
	\bar{E}_i=\beta_i\sum_{k=1}^N\bigg|\frac{{\mathbf{\bar{h}}_k}^H\mathbf{\bar{h}}_i\sqrt{p_k}}{||\mathbf{\bar{h}}_k||}\bigg|^2=\beta_i\sum_{k=1}^N Q_{k,i}p_k,\label{Eilos2}
	\end{align}
	where $Q_{k,i}= {\big|\mathbf{\bar{h}}_k^{H}\mathbf{\bar{h}}_i\big|^2}\big/{||\mathbf{\bar{h}}_k||^2}$ represents the power contribution at $s_i$ of the signal meant to $s_k$, and notice that $Q_{k,k}=\mathbf{w}_k^T\mathbf{\bar{h}}_k=||\mathbf{\bar{h}}_k||^2$. With the above result, $\mathbf{P1}$ can be re-written as a linear programming (LP) problem as
	\begin{subequations}\label{P3}
		\begin{alignat}{2}
		\mathbf{P3:}\ \ &\underset{\mathbf{p},\ \bar{\xi}}{\mathrm{minimize}}       &\qquad & 
		-\bar{\xi} \label{P3:a}\\
		&\text{subject to} & &  \mathbf{B}\mathbf{Q}^T\mathbf{p}\succeq \bar{\xi}\mathbf{1}_{N\times 1}\label{P3:b}\\
		& & &\ \ \ \mathbf{1}^T\mathbf{p}=1\label{P3:c}\\
		& & &\qquad\ \mathbf{p}\succeq\mathbf{0},\label{P3:d}
		\end{alignat}	
	\end{subequations}
	where $\mathbf{B}=\mathrm{diag}\big([\beta_1,\beta_2,\cdots,\beta_N]\big)$. The problem is now composed of $N+1$ linear constraints and variables, and can be solved efficiently. For instance, if interior point methods are used, solving $\mathbf{P3}$ will take at most $\mathcal{O}(\sqrt{N+1}\log(1/\varepsilon))$ iterations, each one with at most $\mathcal{O}((N+1)^3)$ arithmetic operations \cite{Ye.2011}. This is a considerable complexity reduction compared to the optimum SDP-based solutions described in Section~\ref{problem}. For solving $\mathbf{P3}$, Algorithm~1 is proposed, which is a particularly simple interior-point method implementation based on affine scaling \cite{Ye.2011}, and explained below.

	\begin{algorithm}[t!]
		\caption{Low-Complexity Beamforming}
		\begin{algorithmic}[1] \label{alg1}
			\STATE \textbf{Input:} $\mathbf{Q},\ \mathbf{B},\ \delta\in(0,1)$ and $\varepsilon\in(0,1)$ \label{lin1}
			\STATE Set $\mathbf{A}\!=\!\begin{bmatrix}
			\mathbf{1}_{N\times 1}& -\mathbf{B}\mathbf{Q}^T & \mathbf{I}_{N\times N}\\
			0& \mathbf{1}_{1\times N} &  \mathbf{0}_{1\times N}
			\end{bmatrix}$, $\mathbf{\upmu}\!=\![-1, 0, \cdots, 0]^T$ \label{lin3}
			\STATE Set $\mathbf{p}_0$ using \eqref{p0}, $\bar{\xi}_0=\min\{\bar{E}_i|_{\mathbf{p}_0}\}$ and  $\mathbf{\upnu}=\bar{E}_i|_{\mathbf{p}_0}-\bar{\xi}_0$\label{lin4}
			\STATE Set $\mathbf{z}^{(0)}=[\bar{\xi}_0,\mathbf{p}_0,\mathbf{\upnu}]^T$ and $\tau=0$ \label{lin5}
			\REPEAT \label{lin5v2}
			\STATE $\mathbf{Z}^{(\tau)}\!\!=\!\mathrm{diag}(\mathbf{z}^{(\tau)})$, $\mathbf{\uplambda}^{(\tau)}\!\!=\!\big(\mathbf{A}(\mathbf{Z}^{(\tau)})^2\!\mathbf{A}^T\big)^{-1}\!\!\mathbf{A}(\mathbf{Z}^{(\tau)})^2\mathbf{\upmu}$ \label{lin6}
			\STATE $\mathbf{r}^{(\tau)}=\mathbf{\upmu}-\mathbf{A}^T\mathbf{\uplambda}^{(\tau)}$\label{lin7}
			\STATE $\mathbf{z}^{(\tau+1)}=\mathbf{z}^{(\tau)}-\delta (\mathbf{Z}^{(\tau)})^2\mathbf{r}^{(\tau)}/||\mathbf{Z}^{(\tau)}\mathbf{r}^{(\tau)}||$ \label{lin8}
			\STATE  $\tau:=\tau+1$ \label{lin9}
			\UNTIL{$\mathbf{1}^T\mathbf{Z}^{(\tau-1)}\mathbf{r}^{(\tau-1)}<\varepsilon$ and $\mathbf{r}^{(\tau-1)}\succeq \mathbf{0}$} \label{lin10}
			\STATE \textbf{Output:}  $\{p_i=z_{i+1}^{(\tau)}\}_{i=1,\cdots,N}$ and $\tau$ \label{lin14}\label{lin11}
		\end{algorithmic}
	\end{algorithm}
	First, \textbf{P3} is transformed to the form: $\mathrm{minimize}\ \mathbf{\upmu}^T\mathbf{z}$ subject to $\mathbf{A}\mathbf{z}=\mathbf{b}$ and $\mathbf{z}\succeq \mathbf{0}$, which is done by stacking constraints \eqref{P3:b} and \eqref{P3:c} into a single system of equations with $\mathbf{\upmu}$ and $\mathbf{A}$ given in line \ref{lin3} of Algorithm~1. Note that the variable space $\mathbf{z}$ now groups $\bar{\xi}$, $\mathbf{p}$ and the slack vector $\mathbf{\upnu}\!\succeq\! 0$ as defined and initialized in line~\ref{lin5}, while $\mathbf{b}=[\mathbf{0}_{1\times N}, 1]^T$. 
	Note that for initialization,  $\mathbf{p}$ is computed in line~\ref{lin4} according to \eqref{p0}, which is shown to be optimal under certain special circumstances, thus, it should provide a good initial guess; while $\bar{\xi}$ is set to be the minimum available RF energy associated to the channels' deterministic component when using such power allocation; and $\mathbf{\upnu}$ is the corresponding slack vector. In addition to the initial $\mathbf{z}$, the iteration index $\tau$ is also established in line~\ref{lin5}. Lines~\ref{lin5v2}-\ref{lin10} constitute the core of the affine scaling method. Specifically, lines~\ref{lin6}, \ref{lin7} are for computing the dual estimates, $\mathbf{\uplambda}$, and reduced costs $\mathbf{r}$, while line~\ref{lin8} is the updating step consisting of an affine scaling with coefficient $\delta$. The algorithm stops, i.e., convergence is declared, when no cost remains negative\footnote{In practice, it is usually required to relax this and allow the algorithm to stop even if $\mathbf{r}$ still contains very small negative values. This is to counteract numerical precision errors, hence one should use $\mathbf{r}\succeq - \eta\mathbf{1}_{N\times 1}$ with small $\eta$. In Section~\ref{results}, $\eta=10^{-4}$ is used.} ($\mathbf{r}\succeq\mathbf{0}$) and the variation in the objective function is already inferior to the tolerance error $\varepsilon$. Then, the power allocation returned by Algorithm~1 in line~\ref{lin11} is $\varepsilon-$optimal.
	\subsection{Performance Bounds}\label{sbounds}
	By taking advantage of the fairness of the problem solution, the total power budget, and the upper bound of $Q_{k,i}\le Q_{i,i}=||\mathbf{\bar{h}}_i||^2$, which comes from using the Cauchy–Schwarz inequality, 
	one realizes that $\bar{E}_i$ in \eqref{Eilos2} is upper-bounded by
	\begin{align}
	\bar{E}_i\le \bar{E}_{\mathrm{ub}}= \min\{ \beta_i||\mathbf{\bar{h}}_i||^2\} \mathbf{1}^T\mathbf{p}=\!\min\{ \beta_i||\mathbf{\bar{h}}_i||^2\}.\label{up}
	\end{align}
	Unfortunately, such bound is not attainable unless all devices have the same average channel, which is unlikely to happen in practice, or when the path-loss of a certain device is much larger than that of the others such that its received energy dominates in the beamforming design. Meanwhile, a lower bound can be derived by noticing that all entries, especially the non-diagonal entries, of matrix $\mathbf{Q}$ are non-negative. Then, consider the extreme case $\mathbf{Q}=\mathrm{diag}\big(\big\{||\mathbf{\bar{h}}_i||^2\big\}_{\forall i}\big)$ under which $\mathbf{P3}$ can be easily solved to obtain
	\begin{align}
	p_i^*&=\frac{1/(\beta_iQ_{i,i})}{\sum_{k=1}^N 1/(\beta_kQ_{k,k})}=\frac{1}{1+\sum\limits_{k\ne i}\frac{||\mathbf{\bar{h}}_i||^2}{||\mathbf{\bar{h}}_k||^2}\frac{1}{\beta_k}},\label{p0}\\ 
	\bar{E}_i&\ge \bar{E}_{\mathrm{lb}}= \frac{1}{\sum_{k=1}^N\beta_k^{-1}||\mathbf{\bar{h}}_k||^{-2}}.\label{lo}
	\end{align}
	
	In general, $||\mathbf{h}_i||^2$ increases linearly with $M$. Hence, both bounds, \eqref{lo} and \eqref{up}, grow linearly and unbounded\footnote{Note that such analytical results do not violate  the law of conservation of energy in practice due to the more substantial path-loss as compared to beamforming gain.} with $M$, and consequently it is possible to conclude that the actual $\bar{E}$ also grows with $M$.
	Additionally, note that $\bar{E}_\mathrm{ub}\le N \bar{E}_\mathrm{lb}$ due to the inequality between the harmonic mean and the minimum function (i.e., $\min \{v_i\}\! \leq\! N/\sum_{k=1}^N\!\!v_k^{-1}$). Therefore, the gap between $\bar{E}_\mathrm{ub}$ and $\bar{E}_\mathrm{lb}$ is limited by the number of EH devices.
	
	According to our discussions in Section~\ref{statistical}, the lower bound for $\bar{E}_i$ in \eqref{lo},  serves also as a lower bound for the actual $\mathbb{E}[E_i]$. Meanwhile, as the spatial correlation (positively) increases, the average energy actually delivered  may be much larger than that predicted by \eqref{lo}, and even reach (or surpass) the upper bound provided in \eqref{up}. Finally, as $M$ grows larger, not only the proposed beamforming scheme is able to increase the average delivered energy, 
	but it also converges faster. This happens because when $M$ increases, Algorithm~\ref{alg1}'s solution gets closer to the initial guess; in fact, by increasing $M$, $\mathbf{Q}$'s off-diagonal elements decrease, and  $\mathbf{Q}$ tends to asymptotically ($M\rightarrow\infty$) mimic a diagonal matrix (due to the channel-hardening effect), for which \eqref{p0} is optimal power allocation.
	\subsection{Analysis under Rician Fading Channels}\label{Rician}
	In this subsection, Rician fading channels are considered with different LOS factors
	$\kappa_i \ge 0$ \cite[Ch. 2]{Proakis.2001}.
	Under such fading, $\mathbf{\bar{h}}_i$ corresponds to the LOS component of the channel, while $\mathbf{\hat{h}}_i$ represents the scattering contribution. The PB is equipped with a half-wavelength uniform linear array and the signals from all antennas are assumed to experience the same average path-loss.
	Then, $\mathbf{\bar{h}}_i=\sqrt{\frac{\kappa_i}{1+\kappa_i}}e^{\mathbbm{i}\mathbf{\upphi}_i}$ such that $\mathbf{\upphi}_i$  is what $s_i$ observes as the mean phase shift vector among the PB's antenna elements, and $\mathbf{\hat{h}}_i\sim \sqrt{\frac{1}{1\!+\!\kappa_i}}\mathcal{CN}(\mathbf{0},\mathbf{I})$, which accounts for uncorrelated channel components. 
	
	Based on the above assumptions, it holds that $||\mathbf{\bar{h}}_i||^2=\frac{\kappa_i}{1+\kappa_i}M$, and consequently \eqref{up} and \eqref{lo} are modified  such that
	\begin{align}
	\frac{M}{\sum_{k=1}^N\frac{\kappa_k+1}{\beta_k\kappa_k}}\le \bar{E}_i\le M\min\Big\{ \frac{\beta_i\kappa_i}{1+\kappa_i}\Big\}.\label{bounds}
	\end{align}
	Finally, note that
	\begin{align}
	\hat{E}_i&=\frac{\beta_i}{1+\kappa_i}\Tr(\mathbf{W}\mathbf{I})=\frac{\beta_i}{1+\kappa_i}\Tr(\mathbf{W})=\frac{\beta_i}{1+\kappa_i} \label{nlos} 
	\end{align}
	since $\mathbf{R}_i=\frac{1}{1+\kappa_i}\mathbf{I}$ and $\Tr(\mathbf{W})=1$. Observe that even under non-LOS conditions, i.e., $\kappa=0$, the proposed beamforming can provide the same energy level as in a single-antenna PB system. Obviously, such energy will be larger when $\kappa$  increases and/or under the effect of some positive spatial correlation.
	\section{Numerical Results}\label{results}
	This section presents numerical results on the performance of the proposed low-complexity average-CSI based EB scheme under the example-case of Rician fading, described in Section~\ref{Rician}. Note that our proposed scheme does not assume any fading distribution knowledge, thus it works for arbitrary channels (as long as the channel mean vectors are known).
	
	Algorithm~1 is run with $\delta=0.9$ and $\varepsilon=10^{-5}$. The performance results under the optimum CSI-based scheme and the optimum average-CSI based scheme, which were described in Section~\ref{problem}, are used as benchmark. The switching antennas ($\mathrm{SA}$) CSI-free scheme proposed in \cite{LopezAlves.2019} is also considered. Notice that under SA, the PB transmits through one antenna at each time such that all its antennas are used during a coherence block, while no CSI is exploited at all. 
	
	The PB is equipped with a uniform linear array (ULA) such that 	$\mathbf{\upphi}_i\!=\!-\![0,1,\cdots,(M\!-\!1)]^T\pi\sin\theta_i$, where $\theta_i$ is the azimuth angle of terminal $s_i$ relative to the boresight of the PB's antenna array \cite[Ch.~5]{Hampton.2014}.
	The EH devices are assumed to be randomly and uniformly distributed around the PB at distances between $1$ and $10$ m, i.e., in an annulus region of around $311\ \mathrm{m}^2$ of area. A log-distance path-loss model with exponent $2.7$ is considered along with a non-distance dependent loss of $16$ dB \cite{Goldsmith.2005}, e.g.,  $\beta_i\!=\!10^{-1.6}\!\times d_i^{-2.7}$, where $d_i$ is the distance between $s_i$ and the PB. Unless stated otherwise, $M=N=8$, and $\kappa=10$ dB is set for all Rician channels involved.
	\begin{figure}[!t]	
		\includegraphics[width=0.41\textwidth,center]{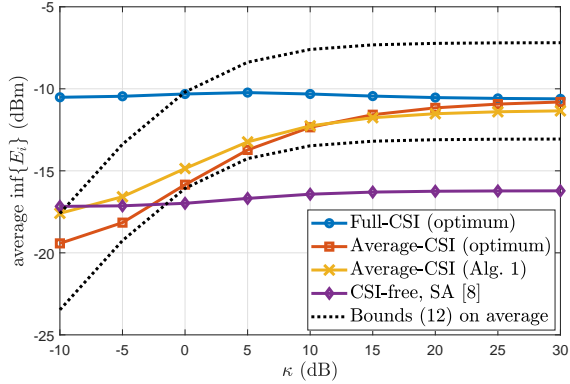}
		\caption{Average worst-case RF energy available at the user devices.} \label{Fig2}
	\end{figure}
	\subsection{Performance Comparison}
	Fig.~\ref{Fig2} corroborates that despite its simplicity, the proposed low-complexity scheme based on average-CSI performs extremely well. In fact, it even outperforms the optimum average-CSI design, which comes from solving the SDP problem  $\mathbf{P2}$, when the Rician factor is below $15$ dB. As the Rician factor increases, the performance gap between the full-CSI and the two average-CSI schemes diminishes, while the CSI-free scheme does not provide additional benefits\footnote{This is without considering the power consumed in the CSI acquisition, which would tilt the scale in favor of the CSI-free and average-CSI schemes.}. In addition, Fig.~\ref{Fig2} validates the bounds given in \eqref{bounds}, and shows that for this particularly scenario, the upper (lower) bound is tighter under small (large) Rician factor.
	
\begin{figure}[!t]
		\includegraphics[width=0.41\textwidth,center]{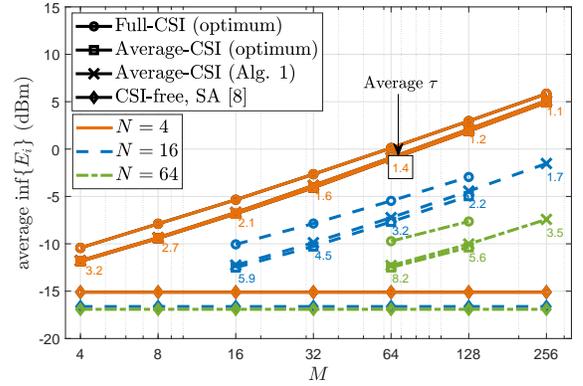}
		\caption{Average worst-case RF energy available at the user devices as a function of the number of PB's antennas for $N\in\{4,16,64$\}.} \label{Fig3}
	\end{figure}

	Fig.~\ref{Fig3} validates the results in Section~\ref{sbounds}, which predicted a nearly linear performance improvement with the number of antennas $M$ at the PB. Notice that as the number of devices increases, the chance of being farther from the PB increases, thus, deteriorating the system performance. Meanwhile, the low-complexity feature of the proposed Algorithm~1 is also evidenced here by showing the average number of iterations that were carried out. As expected, as the number of devices increases, more iterations are required, but less than 10 iterations on average sufficed in all the cases. Increasing the number of antennas is shown to be beneficial in this case as expected from our discussions at the end of Section~\ref{sbounds}.
	Finally, notice that the fast convergence feature of the proposed algorithm contrasts with the SDP-based implementations (both full-CSI and average-CSI) which require considerably more time, and as such it is infeasible to obtain their performance when $M=256$, $N\in\{16,64\}$.
	\subsection{Improvement via PB Antenna Rotation}
\begin{figure}[!t]
		\includegraphics[width=0.41\textwidth,center]{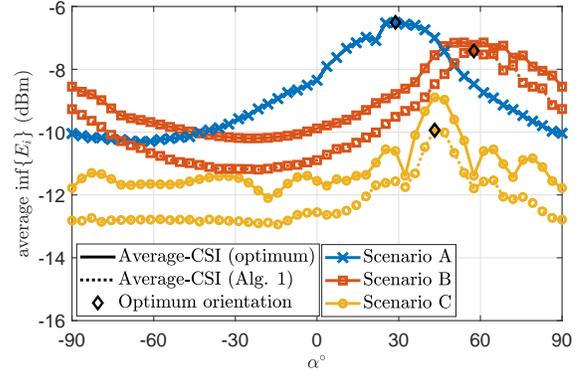}
		\caption{Average worst-case RF energy available at the user devices as a function of the PB's rotation angle for three different setups: i) Scenario A, where $\mathbf{d}=[2, 2, 4, 4, 6, 6, 8, 8]$m and $\theta_i=10i^{\circ}$, ii) Scenario B, where $d_i=(1+i)$ m and $\theta_i=90^{\circ}-10i^{\circ}$, and iii) Scenario C, where $\mathbf{d}=[3, 3, 5, 5, 7, 7, 10, 10]$m and $\mathbf{\uptheta}=[20, 20, 60, 60, 40, 40, 10, 80]^{\circ}$.} \label{Fig4}
	\end{figure}

	The ULA angular orientation directly impacts on $\theta_i$, and consequently on $\mathbf{\upphi}_i$, $\mathbf{\bar{h}}_i$ and $\mathbf{Q}$, influencing the system performance. Assume the PB can adjust its orientation by rotating the array by $\alpha$ radians such that $\theta_i:=\theta_i+\alpha$. This may be possible in static setups, where the task is committed to the technician/user, or in slow-varying environments, where the PB itself is equipped with a rotary-motor. 
	The performance, as a function of such rotation angle, is shown in Fig.~\ref{Fig4} for three different setups. Notice that the antenna array orientation and/or devices' angular position play a major role on the system performance. Intuitively, it would be desirable that the ULA is geared towards the farthest user(s) to counteract the most adverse path-loss(es), however, this is not completely true as evidenced by Fig.~\ref{Fig4}. 
	Also, the actual rotation gains are considerable since the gaps between the global minimums and maximums are around 3 dB for the scenarios shown. Meanwhile, the performance gap between the optimum and the low-complexity design is not greater than 1 dB for all scenarios (0 dB in case of Scenario A), and their curves follow similar trends.
	\section{Conclusion}\label{conclusions}
	This letter presented a low-complexity, yet effective, beamforming scheme that allows a PB to fairly power a set of EH devices. Albeit also applicable with instantaneous CSI, the scheme is proposed and assessed in tandem with the assumption that only first-order statistics of the channels are available, given the context of simple low-power IoT devices. Besides simpler implementation, the proposed scheme outperforms the optimum average-CSI based scheme under low-to-moderate LOS conditions in Rician fading channels, and its performance improves as the number of PB's antennas increases, leveraging on the channel-hardening effect. Also, it was shown that further performance improvement can be obtained via proper angular rotation of the PB.
	Exploring analytically/algorithmically the PB rotation optimization and the performance under different antenna array architectures are future research directions.
	\bibliographystyle{IEEEtran}
	\bibliography{IEEEabrv,references}

\begin{thebibliography}{10}
\providecommand{\url}[1]{#1}
\csname url@samestyle\endcsname
\providecommand{\newblock}{\relax}
\providecommand{\bibinfo}[2]{#2}
\providecommand{\BIBentrySTDinterwordspacing}{\spaceskip=0pt\relax}
\providecommand{\BIBentryALTinterwordstretchfactor}{4}
\providecommand{\BIBentryALTinterwordspacing}{\spaceskip=\fontdimen2\font plus
\BIBentryALTinterwordstretchfactor\fontdimen3\font minus
  \fontdimen4\font\relax}
\providecommand{\BIBforeignlanguage}[2]{{%
\expandafter\ifx\csname l@#1\endcsname\relax
\typeout{** WARNING: IEEEtran.bst: No hyphenation pattern has been}%
\typeout{** loaded for the language `#1'. Using the pattern for}%
\typeout{** the default language instead.}%
\else
\language=\csname l@#1\endcsname
\fi
#2}}
\providecommand{\BIBdecl}{\relax}
\BIBdecl

\bibitem{Lopez.2019}
O.~L.~A. {López \textit{et al.}}, ``Massive wireless energy transfer: Enabling
  sustainable {IoT} towards {6G} era,'' \emph{arXiv preprint arXiv:1912.05322},
  2019.

\bibitem{Zeng.2015}
Y.~{Zeng} and R.~{Zhang}, ``Optimized training design for wireless energy
  transfer,'' \emph{IEEE Trans. Commun.}, vol.~63, no.~2, pp. 536--550, Feb.
  2015.

\bibitem{ZengZhang.2015}
------, ``Optimized training for net energy maximization in multi-antenna
  wireless energy transfer over frequency-selective channel,'' \emph{IEEE
  Trans. Commun.}, vol.~63, no.~6, pp. 2360--2373, Jun. 2015.

\bibitem{Dai.2018}
H.~{Dai \textit{et al.}}, ``Wireless charger placement for directional
  charging,'' \emph{IEEE/ACM Trans. Netw.}, vol.~26, no.~4, pp. 1865--1878,
  Aug. 2018.

\bibitem{Kashyap.2016}
S.~{Kashyap}, E.~{Bj\"ornson}, and E.~G. {Larsson}, ``On the feasibility of
  wireless energy transfer using massive antenna arrays,'' \emph{IEEE Trans.
  Wireless Commun.}, vol.~15, no.~5, pp. 3466--3480, May 2016.

\bibitem{Chu.2018}
Z.~{Chu \textit{et al.}}, ``Wireless powered sensor networks for {Internet of
  Things: Maximum} throughput and optimal power allocation,'' \emph{IEEE
  Internet Things J.}, vol.~5, no.~1, pp. 310--321, 2018.

\bibitem{Clerckx.2018}
B.~{Clerckx} and J.~{Kim}, ``On the beneficial roles of fading and transmit
  diversity in wireless power transfer with nonlinear energy harvesting,''
  \emph{IEEE Trans. Wireless Commun.}, vol.~17, no.~11, pp. 7731--7743, Nov.
  2018.

\bibitem{LopezAlves.2019}
O.~L.~A. {{L\'opez} \textit{et al.}}, ``Statistical analysis of multiple
  antenna strategies for wireless energy transfer,'' \emph{IEEE Trans.
  Commun.}, vol.~67, no.~10, pp. 7245--7262, Oct. 2019.

\bibitem{LopezMontejo.2020}
O.~L.~A. {L{\'o}pez \textit{et al.}}, ``On {CSI}-free multi-antenna schemes for
  massive wireless energy transfer,'' \emph{IEEE Internet Things J.}, pp. 1--1,
  2020.

\bibitem{Thudugalage.2016}
A.~{Thudugalage}, S.~{Atapattu}, and J.~{Evans}, ``Beamformer design for
  wireless\! energy\! transfer\! with\! fairness,'' in \emph{Proc.\! IEEE ICC},
  2016, pp. 1--6.

\bibitem{Ye.2011}
Y.~Ye, \emph{Interior point algorithms: theory and analysis}.\hskip 1em plus
  0.5em minus 0.4em\relax John Wiley \& Sons, 2011, vol.~44.

\bibitem{Proakis.2001}
J.~G. Proakis, \emph{Digital communications}, 4th~ed.\hskip 1em plus 0.5em
  minus 0.4em\relax McGraw-Hill, 2001.

\bibitem{Hampton.2014}
J.~R. Hampton, \emph{Introduction to MIMO communications}.\hskip 1em plus 0.5em
  minus 0.4em\relax Cambridge Univ. Press, 2014.

\bibitem{Goldsmith.2005}
A.~Goldsmith, \emph{Wireless communications}.\hskip 1em plus 0.5em minus
  0.4em\relax Cambridge Univ. Press, 2005.

\end{thebibliography}
\end{document}